# Nonlinear Network Dynamics on Earthquake Fault Systems

by

Paul B. Rundle[1], John B. Rundle[2], Kristy F. Tiampo[3],

Jorge de sa Martins[3],Seth McGinnis[3], and W. Klein[4]

[1]Fairview High School, Boulder CO  80309
(Now at:  Department of Physics
301 E 12th St., Harvey Mudd College,
Claremont, CA 91711)
[2]Colorado Center for Chaos & Complexity
CIRES, and Department of Physics, CB 216
University of Colorado, Boulder, CO 80309
and
Distinguished Visiting Scientist, Jet Propulsion Laboratory
Pasadena, CA 91125
[3]Colorado Center for Chaos & Complexity
and CIRES, CB 216
University of Colorado, Boulder, CO 80309

[4]Department of Physics, Boston University
Boston, MA 02215
and
Center for Nonlinear Science
Los Alamos National Laboratory, Los Alamos, NM 87545

## Abstract

Earthquake faults occur in networks that have dynamical modes not displayed by single isolated faults.  Using simulations of the network of strike-slip faults in southern California, we find that the physics depends critically on both the interactions among the faults, which are determined by the geometry of the fault  network, as well as on the stress dissipation properties of the nonlinear frictional physics, similar to the dynamics of integrate-and-fire neural networks.

Earthquake forecasting and prediction is complicated by the fact that earthquake faults in nature occur in strongly interacting fault networks. To date however, recent work has focussed primarily on models for single isolated faults. Yet it is likely that emergent modes may appear in complex fault networks that are not properties of single faults [1,2]. Such "network modes" [3-9] in nature include enhanced seismic triggering, retardation, temporary quasi-periodic behavior such as observed at Parkfield, California, "Mogi donuts", precursory quiescence or activation, and clustering. Qualitatively similar phenomena are seen in integrate-and-fire neural networks, where complex dynamical patterns arise through the interactions of simple voltage-threshold neural cells [1,2,10]. Here we examine the dynamics of the geometrically complex network of horizontally-slipping strike-slip faults existing in southern California to develop clues for understanding the failure modes characterizing interacting fault networks.

To summarize our results: We found that strongly correlated, geometrically complex mean field fault networks have dynamics very different from single isolated faults. We also find that the stress-dissipation properties of the fault friction law plays an important role in the on-off switching of dynamical activity on the network, as well as in the nature and configuration of the failure modes. The dynamics can be illuminated by the construction of Coulomb Failure Functions, illustrating the influence of one fault on another.

General methods for carrying out the network simulations have been discussed in refs. [8,11,12]. Briefly, one defines a fault geometry in an elastic medium, computes the stress Greens functions (i.e., stress transfer coefficients), assigns frictional properties to each fault, then drives the system via the slip deficit (defined below). The elastic interactions produce mean field dynamics in the simulations [8]. We focus here on the major horizontally-slipping strike-slip (horizontal motion) faults in southern California that produce the most frequent and largest magnitude events. We used the tabulation of strike slip faults and fault properties as published in ref [13]. All major faults in southern California, together with the major historic earthquakes, are shown in figure 1. Figure 2 shows our model fault network. Each fault was assigned a uniform depth of 20 km, the maximum depth of earthquakes in California, and was subdivided into segments having a horizontal scale size of approximately 10 km each.

Several friction laws are described in the literature, including Coulomb failure [14], slip-dependent or velocity-dependent friction [4,14], and rate-and-state [15]. Here we use a parametrization of recent laboratory friction experiments [16], in which the stiffness of the loading machine is low enough to allow for unstable stick-slip when a failure threshold $\sigma^F(V)$ is reached, where $\sigma^F(V)$ is a weak (logarithmic) function of the load point velocity V. Sudden slip then occurs in which the stress decreases to the level of a residual stress $\sigma^R(V)$, again a weak function of V. Stable precursory slip is observed to occur whose velocity increases with stress level, reaching a magnitude of a few percent of the driving load point velocity just prior to failure at $\sigma = \sigma^F(V)$.

The simplest form of the friction equations describing these experiments can be obtained from space-time coarse-graining procedures applied to simple planar fault models [17]. For a single block sliding on a frictional surface, the mean field form of these equations reduces to:

$$\frac{ds}{dt} = \frac{\Delta\sigma}{K}\left\{\alpha + \delta\left(t - t_F\right)\right\} \qquad \text{(Friction Stress)} \qquad (1)$$

$$\sigma = K\left(Vt - s\right) \qquad \text{(Elastic Load Stress)} \qquad (2)$$

where $s(\mathbf{x},t)$ is slip at position $\mathbf{x}$ and time t, $\sigma(\mathbf{x},t)$ is shear stress, K is a elastic stiffness (change in stress per unit slip), and stress drop $\Delta\sigma = \sigma - \sigma^R(V)$. For laboratory experiments, K is the {machine + sample} stiffness, and for simulations, represents the stiffness of a coarse-grained element of the fault of scale size L. $\alpha$ represents the ratio of the effective stiffness modulus KL to a surface viscosity $\eta$, $\alpha = KL / \eta$. $\delta()$ is the Dirac delta, and $t_F$ is any time at which $\sigma(\mathbf{x},t_F) = \sigma^F(V)$. The quantity $(s - Vt)$ is the slip deficit referred to above. Here we examine models with $\alpha = $ constant. Both $\sigma^F$ and $\sigma^R$ can be parametrized as functions of the normal stress $\chi$ by means of coefficients of static $\mu_S$ and ("effective") kinetic $\mu_K$ coefficients of friction, $\sigma^F = \mu_S \chi$, $\sigma^R = \mu_K \chi$. The latter relation implicitly assumes that dynamical overshoot or undershoot during sliding is approximately constant [14].

An interesting result with important implications can be obtained by considering two sliding blocks, coupled to one another by a spring with constant $K_C$, and each coupled to a loader plate by a spring with constant $K_L$. Consider a set of reduced equations, which for block 1 are:

$$\frac{ds_1}{dt} = \frac{\Delta\sigma_1}{\left(K_L + K_C\right)}\alpha \qquad (3)$$

$$\sigma_1 = K_L\left(Vt - s_1\right) + K_C\left(s_2 - s_1\right) \qquad (4)$$

An analogous set of equations holds for block 2. If the difference in stress at time t=0 is given by $\delta\sigma(0) = \sigma_2(0) - \sigma_1(0)$, then at a time t later:

$$\delta\sigma(t) = \delta\sigma(0)\, e^{-\beta t} \quad , \quad \text{where} \quad \beta = \alpha\left(\frac{K_L + 2K_C}{K_L + K_C}\right) \qquad (5)$$

For $\alpha > 0$, differences in stress decay exponentially in time, a process of *stress smoothing*. If either $\alpha < 0$ or that $K_C < 0$, conditions that can occur in more general elastic or frictional systems [17, 18, 19], variations in stress grow exponentially in time, a process of *stress roughening*. For general three-dimensional fault network models, both stress smoothing and stress roughening should occur [18], as well as smoothing-to-roughening transitions.

Earthquake data obtained from the historical record as well as geological field studies represent the primary physical signatures of how the earthquake cycle is affected by the frictional properties that exist on the faults. The timing, magnitude and complexity of these historical events are a direct reflection of the values of the frictional parameters: $\alpha$, $\sigma^F$, $\sigma^R$. Since the dynamics (1)-(2) depends on the characteristic length scale L for each segment, all of these frictional parameters should be regarded as scale-dependent functions of L: $\alpha = \alpha(\Delta\sigma, L)$, $\sigma^F = \sigma^F(L, V)$, $\sigma^R = \sigma^R(L, V)$. For simulations in which one or more distinct scales L are chosen for each fault segment (length and width, for example), one must choose $\alpha$, $\sigma^F$, $\sigma^R$ in such a way that the historical record of activity on the fault network is matched as closely as possible. This is the *data assimilation* problem for which we have developed a simple, but physically motivated method.

For historical earthquakes, there can be considerable uncertainty about where the event was located [4]. Modern studies [4,19,20] of earthquakes indicate that slip or seismic moment $M_o$ ([4]: also defined in (4) below) is often distributed regionally over a number of faults and sub-faults. Therefore our technique assigns a weighted average of the scalar seismic moments for given historic or pre-historic events during an observational period to *all* of the faults in the system. To be physically plausible, the weighting scheme should assign most of the moment $M_o$ to faults near the location of maximum ground shaking and decay rapidly with distance. Since the seismic moment is the torque associated with one of the moment tensor stress-traction double couples, it is most reasonable to use the (inverse cube power of distance r) law that describes the decay of stress with distance [21]. Comparisons with data indicate that this method yields average recurrence intervals similar to those found in nature.

*Step 1: Assignment of Moment Rates.* All historical events in southern California since 1812 are used (ref [13]). For each of the 215 fault segments in the model, the contribution of moment release rate from the $j^{th}$ historical earthquake $dM_o(t_j)/dt$ to the rate on the $i^{th}$ fault segment, $dm_i/dt$, is :

$$\frac{dm_i}{dt} = \Gamma \left[ \frac{\sum\limits_j \frac{dM_o(t_j)}{dt} \; r_{ij}^{-3}}{\sum\limits_j r_{ij}^{-3}} \right] \qquad (6)$$

where $r_{ij} = | \mathbf{x}_i - \mathbf{x}_j |$ is the distance between the event at location $\mathbf{x}_j$ and time $t_j$, and the fault segment at $\mathbf{x}_i$. The factor $\Gamma$ accounts for the limited period of historical data available compared to the length of the earthquake cycle, and is determined by matching the total regional moment rate, $\Sigma_i \; dm_i/dt$, to the observed current regional moment rate. We find $\Gamma \approx .44$. Application of (6) when $\mathbf{x}_i \approx \mathbf{x}_j$ is understood to be in the limiting sense. Equation (6) arises if one regards $r_{ij}^{-3}$ as a probability density function, and assumes that each earthquake is a point source. We correct for the largest events, which are long compared to the depth, by representing the large event as a summation of smaller events distributed along the fault.

*Step 2: Determination of Friction Coefficients.* The seismic moment is:

$$M_o(t_j) = \mu \left\langle s(t_j) \right\rangle A \qquad (7)$$

where $\mu$ is shear modulus, $\langle s(t_j) \rangle$ is average slip at time $t_j$, and $A$ is fault area. For a compact fault, the average slip in terms of stress drop $\Delta\sigma$ is [22]:

$$\left\langle s \right\rangle = \frac{f \; \Delta\sigma \; \sqrt{A}}{\mu} \qquad (8)$$

where $f$ is a dimensionless fault segment shape factor having a value typically near 1. Standard assumptions of $f \sim 1$, $\Delta\sigma \sim 5 \times 10^6$ Pa, $\mu \sim 3 \times 10^{10}$ Pa yield reasonable slip values. The average slip is converted to a difference between static and kinetic friction, $(\mu_S - \mu_K)_i$ for the $i^{th}$ fault segment via the relation:

$$\left( \mu_s - \mu_k \right)_i \approx \frac{m_i}{f \; A^{3/2} \; \chi_i} \qquad (9)$$

obtained by combining (7), (8), and $\Delta\sigma \approx \sigma^F - \sigma^R$. To compute $(\mu_S - \mu_K)_i$, a typical value of $\chi_i$ for each segment is computed from the average gravitationally-induced compressive stress. Since the stochastic nature of the dynamics depends only on the differences $(\mu_S - \mu_K)_i$ [8,11], we set $\mu_k = .001$ (all i).

*Step 3. Aseismic Slip Factor:* Earthquake faults are characterized by varying amounts of aseismic creep (slow slip generating no elastic waves) that arises from the "stress leakage" factor $\alpha$. Analogously in a neural network, $\alpha$ represents the current leakage through the cell membrane. The most famous example of aseismic creep is the region of the San Andreas fault in central California, in which no seismic slip has ever been observed. The average fraction of slip on each fault that is aseismic is equal to $\alpha_i./2$, and has also been tabulated for southern California faults in ref [13].

Figures 3 and 4 are illustrations of the space-time behavior of the Coulomb Failure Function $CFF(\mathbf{x}_i,t) = \{\sigma(\mathbf{x}_i,t) - \mu_S(\mathbf{x}_i) \chi(\mathbf{x}_i,t)\}$ for all of the 215 fault segments, placed end-to-end along the horizontal axis, in the southern California network model (see figure 2). A horizontal line represents an earthquake, which occurs on a segment when $CFF(\mathbf{x}_i,t) = 0$. In figure 3, values for $\alpha_i$ have been assigned using the data in ref [13], whereas in figure 4, all $\alpha_i = 0$.

From figures 3 and 4, it can be seen that changing the values of $\alpha_i$ has a profound effect on the network dynamics. For larger $\alpha_i$ (Figure 3) and excitatory interactions, the stress field is increasingly smoothed and the earthquakes tend to be larger ("decreasing complexity"). For smaller $\alpha_i$ or even inhibiting interactions, the stress field tends to roughen (Figure 4) and the corresponding events are smaller ("increasing complexity"). With smaller $\alpha_i$ (Figure 4), the various fault segments tend to behave more independently than for larger $\alpha_i$ (Figure 3). One can speak of a "roughness length" for the stress field similar in many respects to a correlation length [23, 24]. We predict that physical manifestations of friction laws on faults are revealed by the space-time patterns in the network dynamics. Dynamical switching of activity due to fault interactions should also be observed. An example is shown in figure 3, in which the south-central region of the San Andreas (left arrow) tends to switch off activity on the eastern Garlock fault (right arrow). Dynamical switching of activity may have already been revealed through observations of real fault networks [3,9]. We are not at present aware of any existing observations in nature relating to stress smoothing or stress roughening transitions. Similar smoothing and roughening processes for the cellular potential should be present in neural networks due to the current leakage through the cell membrane.

Acknowledgments. Research by PBR was supported by the Southern California Earthquake Center (Contribution 540) under NSF grant EAR-8920136. Research by JBR was funded by USDOE/OBES grant DE-FG03-95ER14499 (theory), and by NASA grant NAG5-5168 (simulations). Research by WK was supported by USDOE/OBES grant DE-FG02-95ER14498. KFT and SM were funded by NGT5-30025 (NASA), and JdsM was supported as a Visiting Fellow by CIRES/NOAA, University of Colorado at Boulder. We would also like to acknowledge generous support by the Maui High Performance Computing Center, project number UNIVY-0314-U00.

**Figure Captions.**

Figure 1. Map of the major faults (blue lines) in southern California, together with largest historic earthquakes (colored circles) occuring since 1812 (from http://www.scecdc.scec.org/clickmap.html/)

Figure 2. Major strike-slip faults in southern California used in themodel. Relative scaled values of friction difference ($\mu_S - \mu_K$) are shown superposed above each fault. (Fault Key: SA, San Andreas; SJ, San Jacinto; ELS, Elsinore; IV, Imperial Valley; LS, Laguna Salada; GAR, Garlock; PV, Palos Verdes; SC, Santa Cruz Island; PISG, Pisgah; BR, Brawley; SM, Santa Monica; LAN, Landers)

Figure 3. Plot of Coulomb Failure Functions plotted as a function of time vs spatial location for 2000 simulation years for a set of values $\{\alpha_i\}$ obtained by approximately matching the ratio of aseismic creep to seismic slip on various fault segments from tabulated data in ref [13]. We actually plot Log$\{$1 - CFF($\mathbf{x}$,t) $\}$ so that subtle differences can be seen. Low CFF($\mathbf{x}$,t)'s are represented by cool colors, high CFF($\mathbf{x}$,t)'s near 0 are represented by hot colors.

Figure 4. Same as figure 3 but with all $\alpha_i$ = 0.

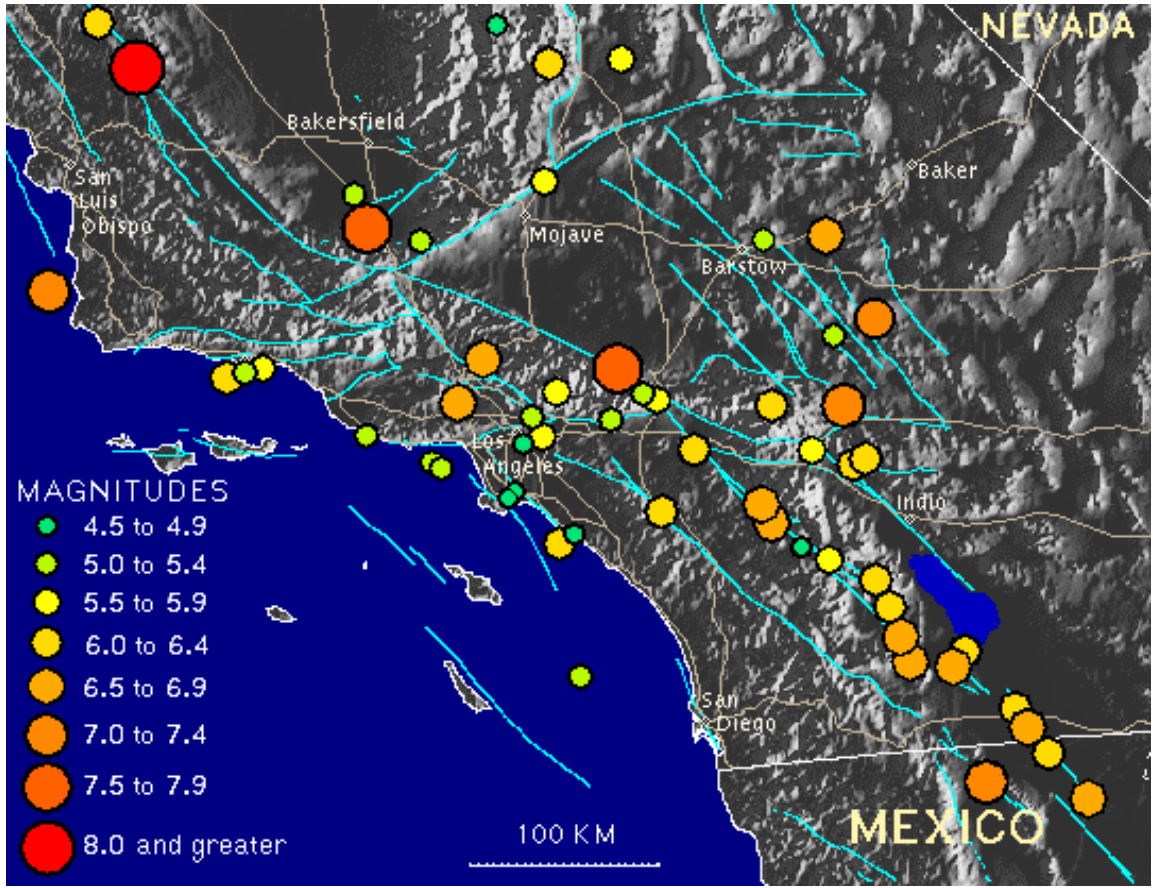

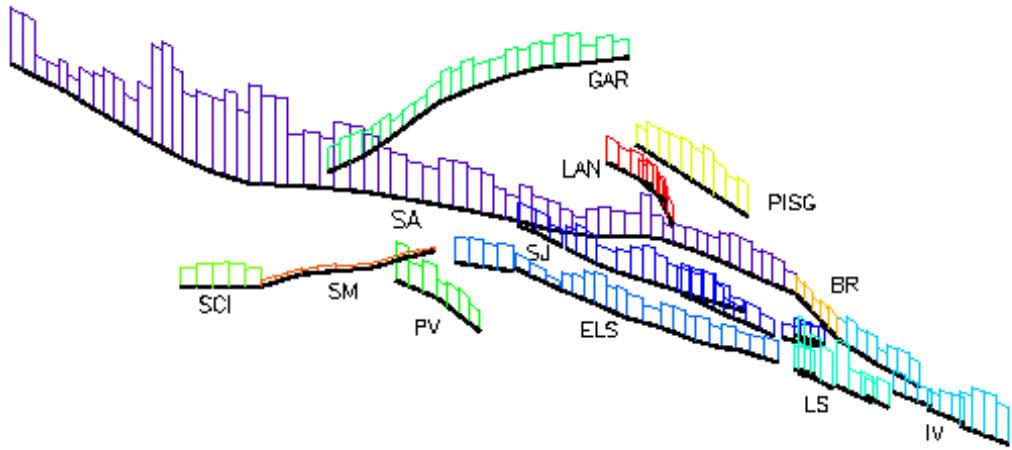

Color—Coded Fault—Friction Map

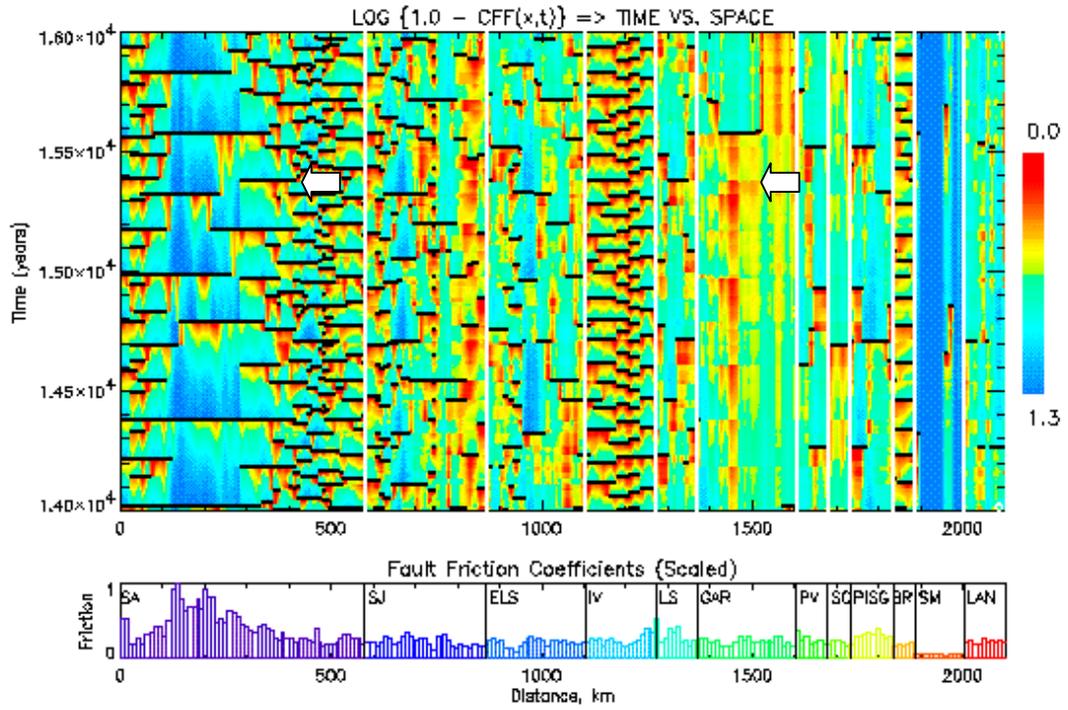

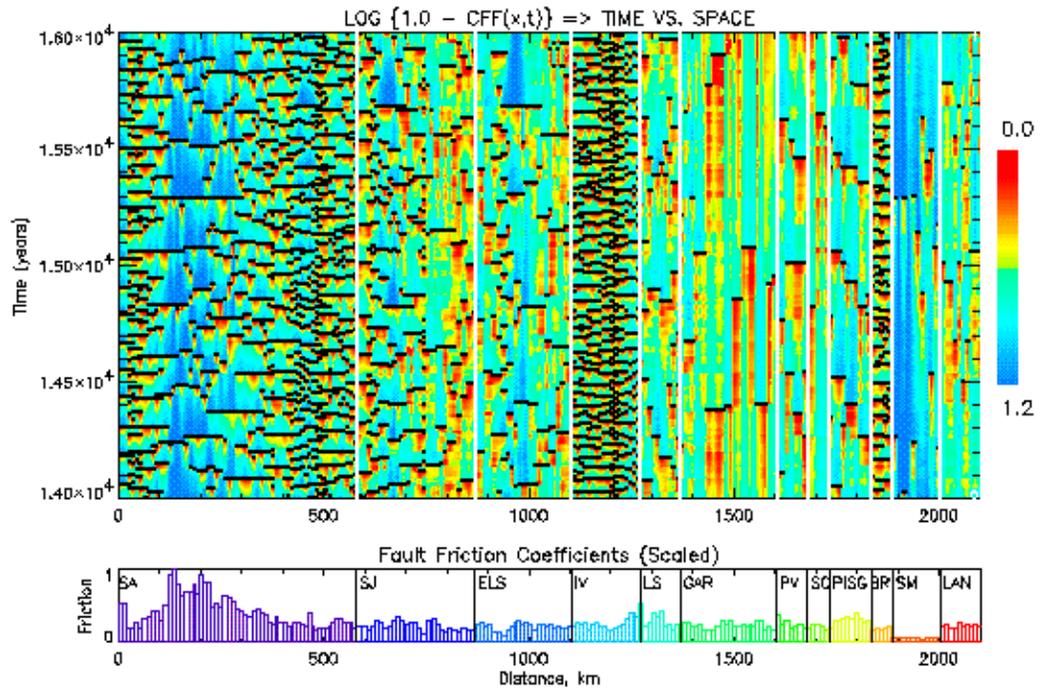